\documentclass[preprint2]{emulateapj}
\usepackage{amsmath}
\usepackage{natbib}
\usepackage{graphicx}
\usepackage{epsf}
\usepackage{color}
\usepackage{threeparttable}
\usepackage{epsfig}
\DeclareGraphicsExtensions{.jpg,.pdf,.png,.eps,.ps}

\newcommand{\ltsima}{$\; \buildrel < \over \sim \;$}
\newcommand{\ltsim}{\lower.5ex\hbox{\ltsima}}

\begin{document}

\title{On the Ratio of Circumference to Diameter for the Largest
  Observable Circles:  An Empirical Approach}

\author{L. Knox}

\affil{Department of Physics, University of California,
Davis, CA 95616, USA}

\email{lknox@ucdavis.edu}

\begin{abstract}
  I present here a measurement of $\pi$ as determined for the largest
  observable circles.  Intriguingly, the value of 16/5 asserted by the
  House of Representatives of the State of Indiana in 1897 is still
  viable, although strongly disfavored relative to 22/7, another
  popular value.  The oft-used `small-circle' value of 3 is ruled out
  at greater than 5$\sigma$.  We discuss connections with string
  theory, sterile neutrinos, and possibilities for (very large) lower
  limits to the size of the Universe.
  \end{abstract}


\bigskip\bigskip


\section{Introduction}

The ratio of the circumference of a circle to its diameter is commonly
referred to as $\pi$ and is one of the most important constants, with
many applications across a great many fields of science.  It is the
only mathematical constant that has its own widely-recognized day of
celebration\footnote{That day is, of course, 3/14.  In Noth America
  this is March 14th, while in much of the rest of the world the
  interpretation is not quite as straightforward but usually taken to
  be the 3rd of February using the convention that months numbered
  greater than 12 are to be understood mod 12.}.  

The value of $\pi$ has been controversial at least since the time
Pythagoras drowned a man for asserting its irrationality.  This
controversy persists despite theoretical calculation of its value to
ridiculously high precision and proofs of its irrationality.

Given the importance of $\pi$ it is surprising that there is not much of a literature on its measurement since Gauss effectively measured it by testing that the sums of the interior angles of a triangle formed by 3 mountain tops added up to 180 degrees\footnote{Zoltan Haiman, personal communication}.  I provide a measurement here.

This work is inspired by the Indiana House of Representatives'
unanimous vote for a bill that asserted the value of $\pi$ as 3.2.
Since this was an astablishment of the value of $\pi$ by legislative
means we will refer to it as $\pi_{Leg}$.  Most people react with
astonishment to this vote, immediately jumping to the conclusion that
it is self-evidently false.  Even at the time the Senate did not go
along, so it never became law.  But what if the basis for the House
Bill, and indeed its actual meaning, has been misunderstood all this
time?  What if they were on to something?

It is especially troubling that the Senate did not even allow a
vote on the measure.  How can we know the truth when this bill
never even received the open give and take of parliamentary
debate?  Was the Senate trying to hide something?

Fortunately, there are other ways to know.  We can {\em measure} it!

An immediate question is what kind of circle we should use. 
Circles around spherical mass distributions have $\pi < \pi_{\rm
  flat}$ where $\pi_{\rm flat}$ is the ratio of circumference to
diameter in Euclidean geometry.  Whereas $\pi_{\rm Leg} > \pi_{\rm
  flat}$.  Such circles will not help us to see how 3.2 can be the
correct value.  

Thus we turn to cosmological circles. The larger a circle we consider, the
greater the effect of curvature and so we wish to study the largest
possible circles.  Here we conservatively limit ourselves to our own 
horizon, though consider larger circles in the Discussion.  

\section{Result}
I assumed a $\Lambda$CDM cosmology with the usual six vanilla
parameters, plus one parameter for the mean curvature.  Using a chain
available on the LAMBDA archive that included WMAP7 \citep{larson11},
BAO \citep{percival10} and $H_0$ measurements \citep{riess09}, as described in
\citet{komatsu11}, I calculated the comoving angular-diameter distance
to the horizon, $D_A$ and also the comoving distance to the horizon,
$r_H$.  Then I defined
$$
\pi_{\rm Hor} = 2 \pi_{\rm flat} D_a/(2 r_H)  = \pi_{\rm flat} D_A/r_H.
$$
Histogramming the chain results in the probability distribution
displayed in Fig. 1.

\begin{figure}[t]\centering
\includegraphics[width=0.5\textwidth]{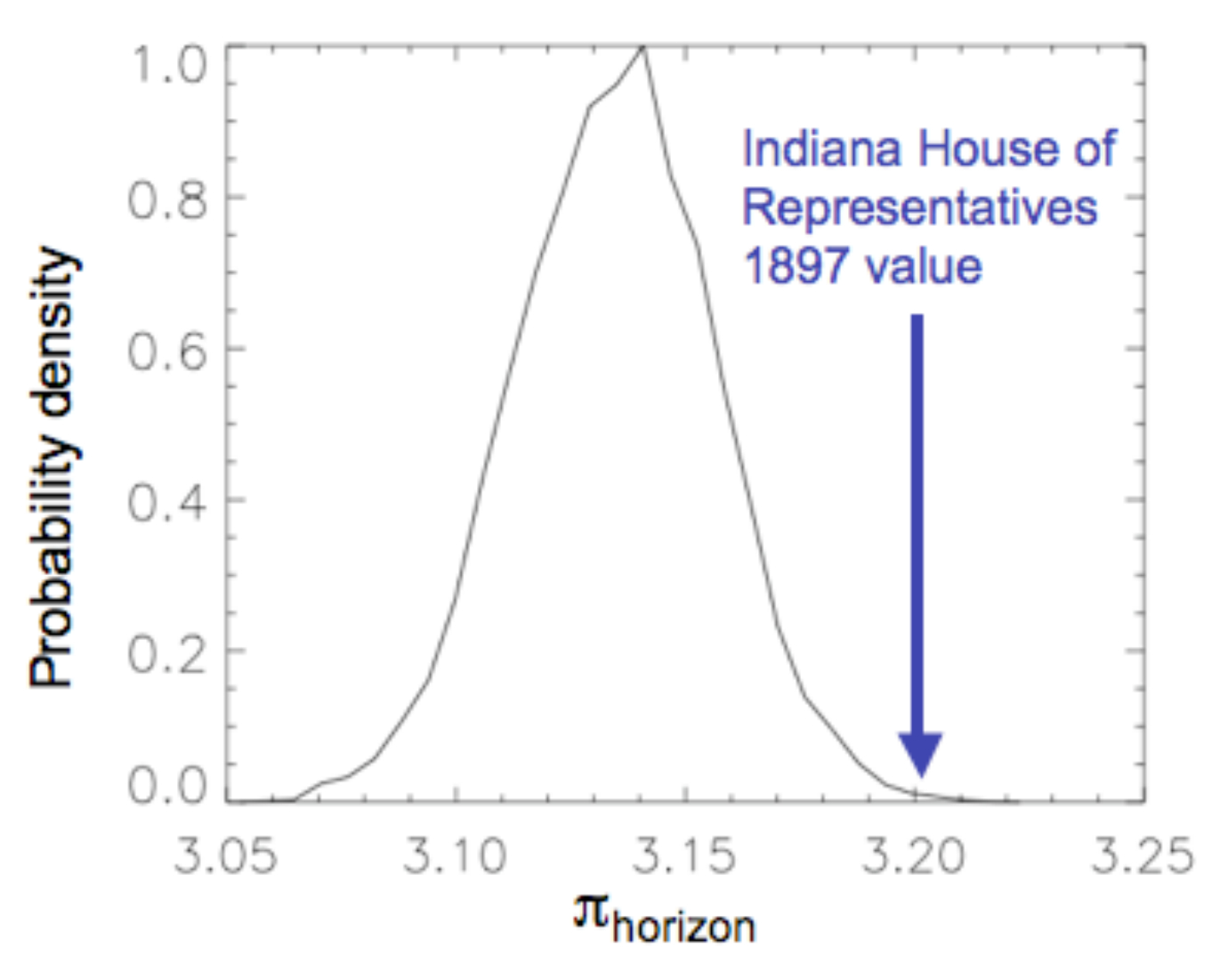}
  \caption[]{Probability distribution for $\pi_{\rm Hor}$ given {\it WMAP}7,
  BAO and $H_0$ data and the assumption of a 7-parameter 
$\Lambda$CDM model.}

\end{figure}

\section{Discussion}

Note that in a $\Lambda$-dominated Universe with a small amount of
mean curvature the value of $\pi$ varies with time.  We define a
``Legislative Universe'' (LU) as one for which $\pi_{\rm Hor}$ = 3.2
at some point in time.  In future
work we will determine the boundary in the $\Omega_K$-$\Omega_V$ plane
dividing LUs from non-LUs.  

We note that LUs require a negative spatial curvature (positive $\Omega_K$).
{\em Perhaps} coincidentally, from the string theory landscape we also
expect our Universe to be one with negative spatial curvature, since
tunneling events result in such a geometry.  Or perhaps this
``coincidence'' points toward some deep connection between string
theory and the theoretical underpinnings of the Legislative value.  
We tentatively posit that while $\pi_{\rm Hor} > \pi_{\rm flat}$ would be
evidence for string theory, $\pi_{\rm Hor} = 3.2$ would be evidence
of this connection.  It might instead be evidence of a connection with the de Sitter
equilibrium cosmology \citep{albrecht11}, for which one also expects a
small negative curvature.  

If one adopts a prior that the Legislative value must hold for {\em some}
circle in the Universe, then we can turn the arguments used here
around to place a lower limit on the size of the Universe.  As long as
mean curvature is less than zero (no matter how close it is to zero), 
if the Universe is big enough, the Legislative value will hold for a big
enough circle.

We have explicitly assumed the standard cosmological
model extended to include non-zero mean curvature.  Departures from
this model would change our interpretation of the data.  In
particular, if we consider a model with {\em fewer} species of neutrinos then this
increases the sound horizon at last scattering \citep{hou11}.  In
order to keep the angular size of the sound horizon fixed to the
observed value, the increased sound horizon means the model would have
to adjust in a way to increase the angular-diameter distance to last
scattering.  One way to do this is to increase  $\Omega_K$
\citep{smith11}.  Since the data prefer not fewer species, but an
excess of species, allowing $N_{\rm eff}$ to vary would put further
pressure on $\pi_{\rm Hor} = \pi_{\rm Leg}$.  Conversely, adopting a
``Legislative prior'' would increase the posterior probability for the
standard model value of $N_{\rm eff}$ relative to $N_{\rm eff}=4$.  

We look forward to new data from the South Pole Telescope and {\it
  Planck}, combined with new BAO measurements, from which we can place
tighter constraints on $\pi_{\rm Hor}$, or, by adopting the Legislative
prior, interesting lower limits to the size of the Univerese.

\begin{acknowledgments}
 I thank Ryan Foley for encouraging the communication
of this work and Zoltan Haiman for further encouragement as the final hours of April 1, 2012
were suddenly upon me.
\end{acknowledgments}


\end{document}